\documentclass[10pt,letterpaper]{article}
\usepackage[top=0.85in,left=2.75in,footskip=0.75in]{geometry}
\usepackage{comment}
\usepackage[normalem]{ulem}

\usepackage{comment}
\usepackage{amsmath,amssymb}
\usepackage{changepage}
\usepackage[utf8x]{inputenc}
\usepackage{textcomp,marvosym}
\usepackage{cite}
\usepackage{nameref,hyperref}
\usepackage[right]{lineno}
\usepackage{lscape}
\usepackage{microtype}
\DisableLigatures[f]{encoding = *, family = * }
\usepackage[table]{xcolor}
\usepackage{array}
\usepackage{subfigure}
\usepackage{multirow}
\newcolumntype{+}{!{\vrule width 2pt}}
\newlength\savedwidth

\newcommand\thickhline{\noalign{\global\savedwidth\arrayrulewidth\global\arrayrulewidth 2pt}%
\hline
\noalign{\global\arrayrulewidth\savedwidth}}
\raggedright
\usepackage[aboveskip=1pt,labelfont=bf,labelsep=period,justification=raggedright,singlelinecheck=off]{caption}

\makeatletter
\renewcommand{\@biblabel}[1]{\quad#1.}
\makeatother
\usepackage{lastpage,fancyhdr,graphicx}
\usepackage{epstopdf}
\fancyheadoffset[L]{2.25in}
\fancyfootoffset[L]{2.25in}
\newcommand*{\inlineequation}[2][]{%
  \begingroup
    \refstepcounter{equation}%
    \ifx\\#1\\%
    \else
      \label{#1}%
    \fi
    \relpenalty=10000 %
    \binoppenalty=10000 %
    \ensuremath{%
      #2%
    }%
  \endgroup
}

\begin{document}
\vspace*{0.2in}

\begin{flushleft}
{\Large \textbf\newline{The multi-channel potentiostat: Development and Evaluation of a Scalable Mini-Potentiostat array for investigating electrochemical reaction mechanisms}}
\newline\\
Pattawong Pansodtee\textsuperscript{1},
John Selberg\textsuperscript{1},
Manping Jia\textsuperscript{1},
Mohammad Jafari\textsuperscript{2},
Harika Dechiraju\textsuperscript{1},
Thomas Thomsen\textsuperscript{1},
Marcella Gomez\textsuperscript{2},
Marco Rolandi\textsuperscript{1},
Mircea Teodorescu\textsuperscript{1},\\

\bigskip
\textbf{1} Department of Electrical and Computer Engineering, University of California Santa Cruz, Santa Cruz, California, United States of America \\
\textbf{2} Department of Applied Mathematics, University of California Santa Cruz, Santa Cruz, California, United States of America \\
\bigskip
\textcurrency Current Address: Department of Electrical and Computer Engineering, University of California Santa Cruz, Santa Cruz, California, United States of America *ppansodt@ucsc.edu
\end{flushleft}


\section*{Abstract}

{

A potentiostat is an essential piece of analytical equipment for studying electrochemical devices and reactions. As the design of electrochemical devices evolve applications for systems with multiple working electrodes have become more common. These applications drive a need for low-cost multi-channel potentiostat systems. We have developed a portable low-cost scalable system with a modular design that can support 8 to 64 channels at a cost as low as \$8 per channel. This design can replace the functionality of commercial potentiostats which cost upwards of \$10k for certain applications. Each channel in the multi-channel potentiostat has an independent adjustable voltage source with a built-in ammeter and switch, making the device flexible for various configurations. The multi-channel potentiostat is designed for low current applications (nA range), but its purpose can change by varying its shunt resistor value. The system can either function as a standalone device or remotely controlled. We demonstrate the functionality of this system for the control of a 24-channel bioelectronic ion pump for open- and closed- loop control of pH.


\section{Introduction}
In the last few decades, the use of miniaturized electrochemical devices has grown rapidly and found diverse applications in scientific and consumer products (e.g., batteries \cite{nitta2015li}, Glucose sensors \cite{strakosas2019non} \cite{Zhaoeaaz0007}, and Breathalyzer \cite{bihar2016disposable}). Unfortunately, the process of developing specialized electrochemical devices is often time-consuming and expensive \cite{abdulbari2017electrochemical}. Figure \ref{fig:basic_diagram}. shows a typical electrochemical experimental setup, where it used measurement equipment such as a potentiostat. A potentiostat is an instrument that controls the voltage between two or more electrodes. It is an essential measurement pieces of equipment used to investigate electrochemical reaction mechanisms using electroanalytical methods \cite{gopinath2005inexpensive}. The main role of the potentiostat is to control the electrochemical reaction using either two electrodes (a working electrode and a counter electrode) or three electrodes (a working electrode, a counter electrode, and a reference electrode). The accuracy and precision of the applied or measured voltages and currents depends on the quality of the electronic hardware, which for commercially available potentiostats, often correlate with the price and portability of each unit. Consequently, one of the challenges faced by the research community is how to test and operate electrochemical devices with less expensive and more portable equipment without compromising the quality of the experiment \cite{adams2019ministat}. Due to cost and access to a quality components, access to quality potentiostats is usually limited, especially in a developing country \cite{rowe2011cheapstat}. As a result, a potentiostat is a significant bottleneck in the process of developing and testing electrochemical devices \cite{cruz2014low}.



\begin{figure}
    \centering
    \includegraphics[width=\linewidth]{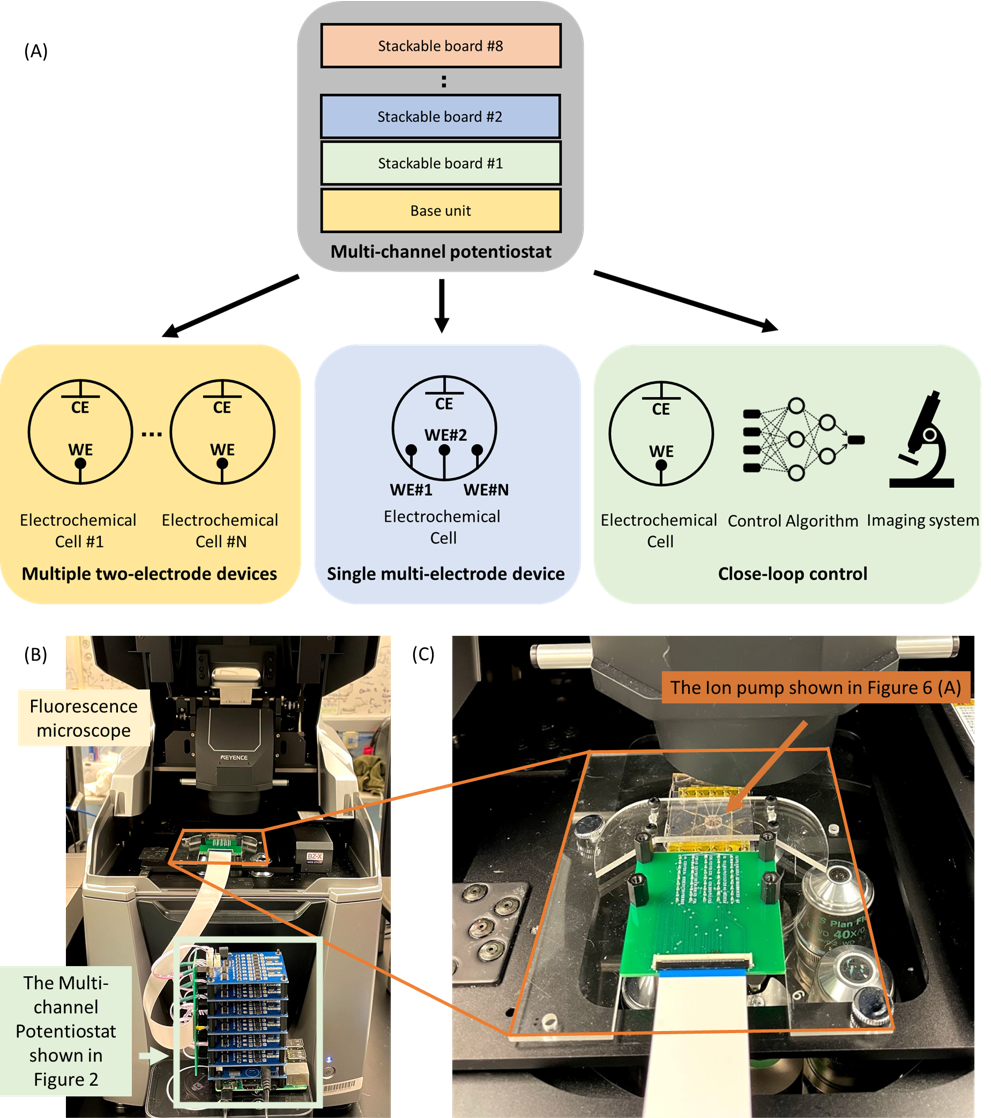}
    \caption{{\bf Multi-channel potentiostat} (A) An example of Multi-channel potentiostat use cases. The multi-channel potentiostat can be setup in various configurations (e.g. multiple two-electrode devices, single multi-electrode devices). (B) Close-loop control experiment with an Ion pump and KEYENCE, BZ-X710 fluorescence microscope. This is also shown in Figure \ref{fig:amperometry:a} (C) Close-up picture of an Ion pump, 9 electrodes device.}
    \label{fig:basic_diagram}
\end{figure}




Portability is a major obstacle in testing small, wearable devices. For instance, large and expensive bench-top potentiostats are not feasible options for operating wearable glucose sensors \cite{adams2018miniature}. Cumyn et al  \cite{cumyn2003design} and Li et al \cite{li2004multi} proposed two multi-channel potentiostats that consist of interconnected independent pieces of off-the-shelf lab  equipment. Consequently, the cost per channel would be relatively high, and the potentiostats would lack portability. While electrochemical devices become more sophisticated and diverse, the number of electrodes and their operating range increases. For instance, the emerging trend of sensor arrays such as Gao et al \cite{gao2016fully}, Zao et al \cite{Zhaoeaaz0007}, and Xu et al \cite{xu2019battery} sweat sensor arrays have more than five electrodes. Selberg et al \cite{selberg2020pump} and Jia et al \cite{jia2020clpump} proposed bioelectronic actuators that have as many as 24 electrodes and require an actuation range of 0-3 V.

Turner et al \cite{turner1987cmos}, Kakerow et al \cite{kakerow1995low}, Stanacevic et al \cite{stanacevic2007vlsi}, Ayers et al \cite{ayers2007design}, and Martin et al \cite{martin2009fully} proposed low-level VLSI (CMOS/single Chip design) single-channel potentiostats. Although, they are more affordable and smaller, they are difficult to replicate. Therefore, there are other approach that utilize commercial off-the-shelf components.
For an instance, Meloni \cite{meloni2016building}, Li et al \cite{li2018easily}, proposed open-source single-channel potentiostats, Jenkins et al \cite{jenkins2019abe} and Ainla et al \cite{ainla2018open} proposed open-source wireless systems, Glasscott et al \cite{glasscott2019sweepstat} designed a Labview controlled device for low-current applications. Rowe et al \cite{rowe2011cheapstat} proposed a high-level circuit board design of a portable less expensive potentiostat (less than \$200) that supports various electroanalysis methods (i.e., cyclic, square wave, linear sweep, and stripping voltammetry) over the voltage range of -0.99 to +0.99V. Friedman et al\cite{friedman2012cost}, Dryden et al \cite{dryden2015dsta}, Cruz et al \cite{cruz2014low}, Hoilett et al \cite{hoilett2020kickstat}, and Adams et al \cite{adams2019ministat} proposed similar devices with smaller form-factor. Dobbelaere et al\cite{dobbelaere2017usb} potentiostat design can reach up to 8V where most of the affordable solutions output range  less than 1 V. Commercial potentiostats \cite{Palmsens:2020, MetrohmAutolab:2013, Nuvan:2011} are not open source, significantly more expensive, and not all of them support multiple electrode setup (e.g., \cite{Palmsens:2020}). Although most of the  potentiostats described here support up to three electrodes, none of them simultaneously address portability, the high number of channels,  the high number of electrodes, the wide output range, and the low-cost. Table \ref{tab:comparison} in Appendix 1 shows a comparison between the proposed system and single channels and multi-channel potentiostats.

This paper reports the design of a potentiostat that is open-source, multi-channel, low-cost, low-current, and scalable. Section 2 explores the proposed multi-channel potentiostat overall architecture, including output, input stages, and the design choices that enable it to feature multiple channels, a wide output range, and low current measurement capability. In section 3, the input and output stages of the multi-channel potentiostat are characterized.
The validation process of the multi-channel potentiostat's cyclic voltammetry ability is to compare the results with
the bench-top Autolab, PGSTAT128N potentiostat on Palladium microelectrode, AgCl electrode in NaCl solution. The system is also used to perform amperometry on an array of electrophoretic ion pumps \cite{selberg2020pump} \cite{jia2020clpump}, to show a correlation between the currents measure and changes in fluorescent intensity.
The demonstration of the multi-channel potentiostat in Section 4. In this section, an external laptop is the running machine learning control algorithm. It performs closed-loop control on electrophoretic ion pumps with feedback from fluorescent imaging, where the machine learning algorithm is presented in further detail in \cite{jafari2020feedback} Section 5 is concludes our report.

\section{Materials and Methods}
\begin{figure}[t]
    \centering
    \subfigure[The assembled system]{\includegraphics[width=.4\textwidth]{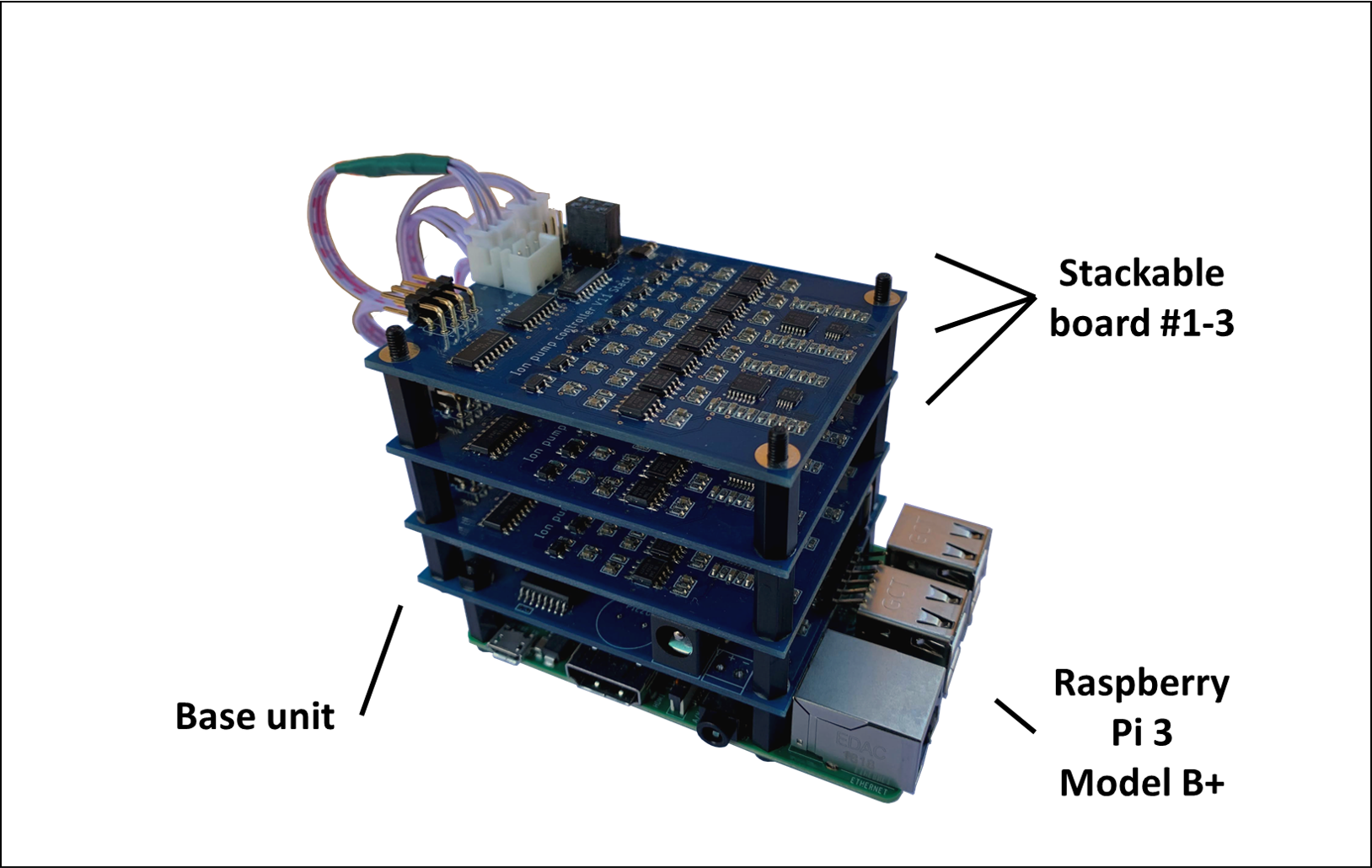}}\\
    \subfigure[Base unit]{\includegraphics[width=.4\textwidth]{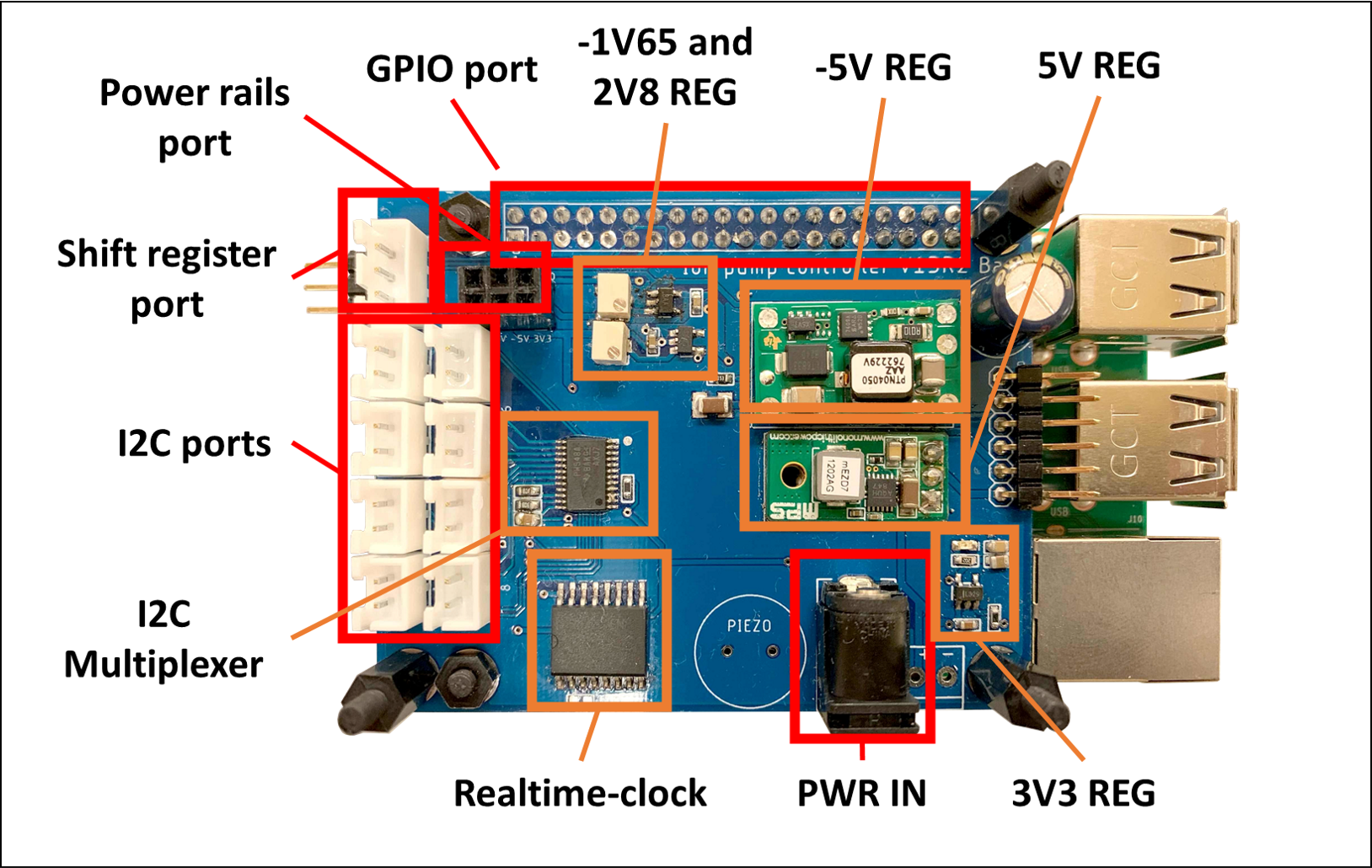}}
    \subfigure[Stackable board]{\includegraphics[width=.4\textwidth]{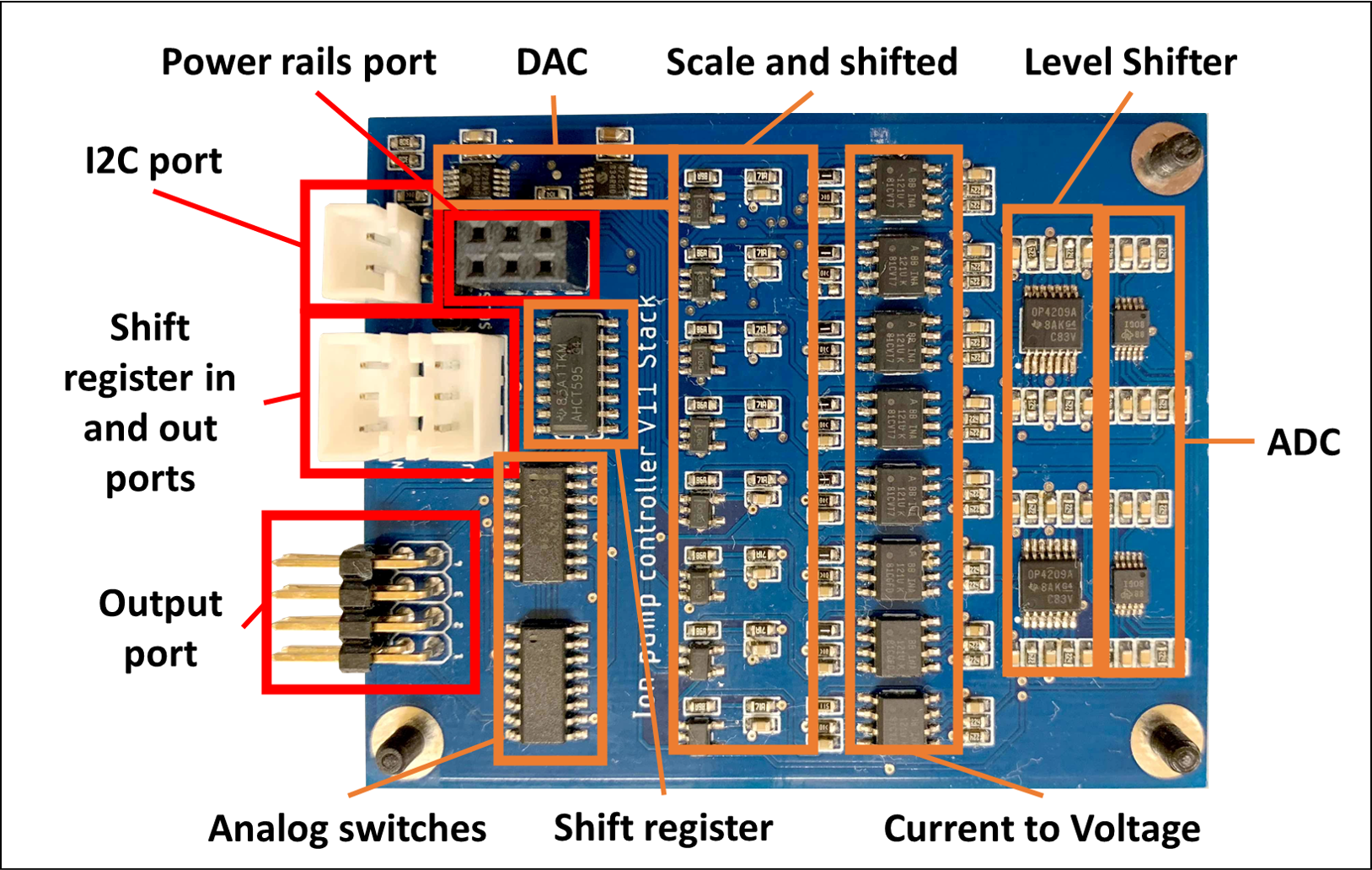}}
    \caption{{\bf The multi-channel potentiostat.} (A) A 24 channels assembled unite with three stackable boards (B) The base unit attached to the Raspberry Pi 3 B+ (C) Stackable board. For (B) and (C), the main electronic Integrated circuit (IC) are highlighted in Orange and connection ports are highlighted in Red.}
    \label{fig:board_pic}
\end{figure}

We are proposing an affordable, scalable, and versatile device that can perform low current cyclic voltammetry and amperometry measurements on various electrode configurations. The proposed design consists of one base unit that attaches to a RaspberryPi 3 B+ board computer and several stackable multi-channel potentiostat boards. Figure \ref{fig:board_pic} shows an assembled unit (a), the base unit attached to the Raspberry Pi 3 B+ (b), and the stackable board (c).

The base unit contains the voltage regulators interfacing with the stackable boards. It consists of a single-board computer (SBC) and a custom Hardware Attached on Top (HAT) printed circuit board (PCB) that provides an interface between the SBC and the stackable boards as well as supply the necessary power. In this paper, the SBC is a low-cost Raspberry Pi 3 Model B+ computer that has Wi-Fi, Ethernet, 40-GPIO connectivity, and runs Raspbian OS version 10 (Buster) and Python 3.5. The multi-channel potentiostat has two operation mode, Standalone (that operate the multi-channel potentiostat independently) and API mode (the multi-channel potentiostat is externally controlled). The HAT is designed in the form-factor of a Raspberry Pi and can accommodate up to eight stackable boards. The HAT contains +5, -5, +3.3, +2.8, and -1.65 V voltage regulators, TCA9548A, I2C multiplexer, DS3231 real-time clock, and a backup battery. It can interface with a Raspberry Pi via GPIO pin headers, one I2C communication bus, and one shift register bus.

Each stackable board contains eight analog potentiostats with adjustable voltage sources, built-in ammeters, analog switches, and digital communication buses to the base unit. Each board is powered by a six-pin stacking header and interfaces with a Base unit via one of a 4-pin I2C cables. A Shift register is either chained with a custom-designed HAT or with another stackable board's port. The most basic configuration consists of one base unit and one stackable board, but the system has a maximum of eight stackable boards for each single base unit. As a result, the multi-channel potentiostat support ranges from a minimum of 8 channels to a maximum of 64 channels. The multi-channel potentiostat has an adjustable voltage source of \(\pm\) 4 V and maximum current of \(\pm\) 1.5 $\mu A$. Therefore, it is used for micro-electrode electrochemistry experiments in both aqueous and non-aqueous, solutions {\cite{bard2001fundamentals}}.

\begin{figure}[h]
    \centering
    \includegraphics[width=\linewidth]{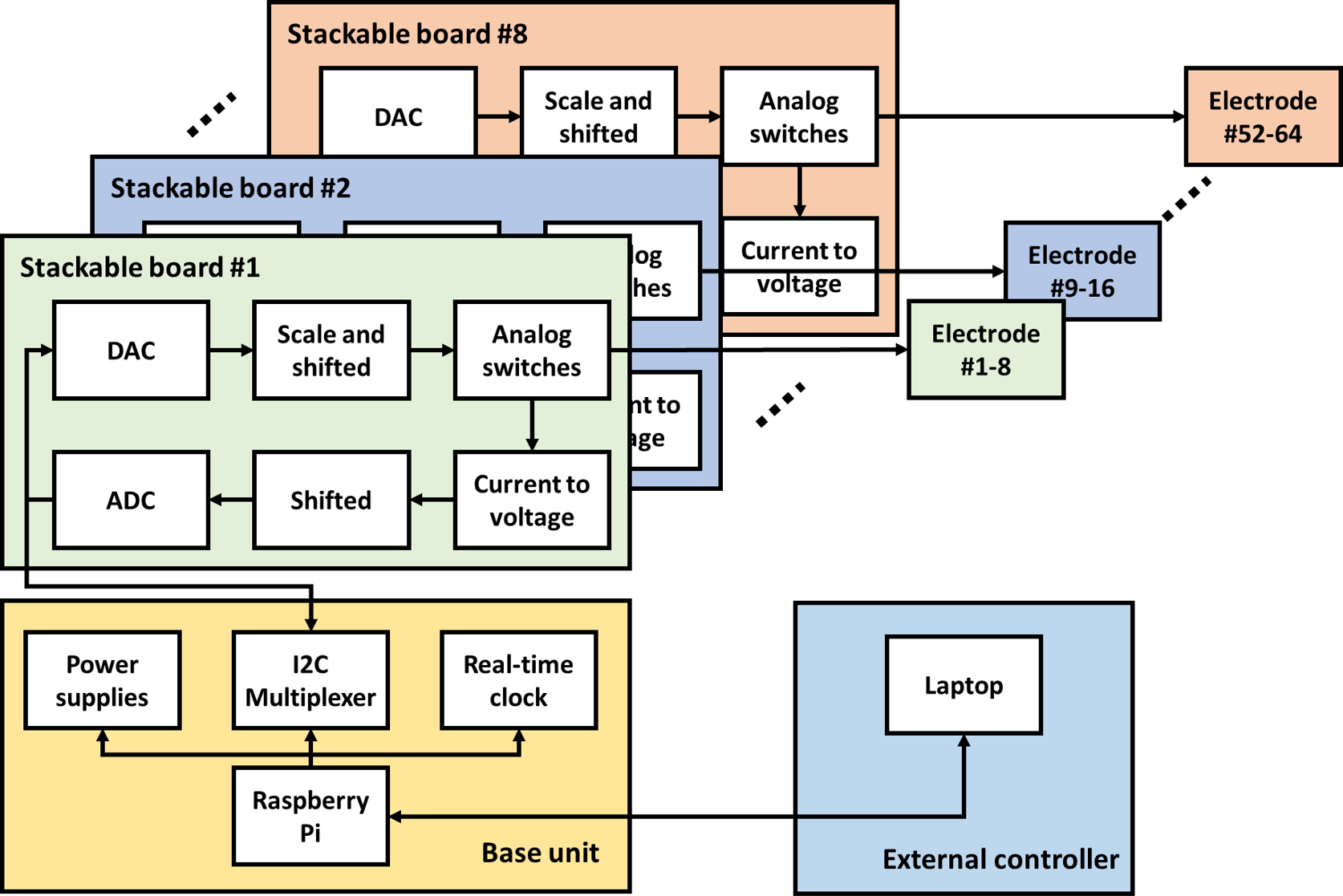}
    \caption{{\bf The multi-channel potentiostat block diagram.} The multi-channel potentiostat consists of a base unit and stackable boards. The base unit receives commands from a laptop and provides power and communication to the stackable boards. Each stackable board contains eight independent channels. The multi-channel potentiostat design allows up to eight stackable boards installed simultaneously, resulting in a total of 64 channels. A computer can interface with the multi-channel potentiostat via Ethernet or Wi-Fi.}
    \label{fig:circuit_diagram}
\end{figure}

The potentiostat circuit is split into two main stages: output and input. The output stage consists of three components: a Digital-to-Analog Converter (DAC), a level-shifter, and an analog switch. The DAC used in the proposed design is Microchip Technology MCP4728, which is a quad buffered 12-bit integrated circuit with a built-in high precision internal voltage reference, programmable I2C address, and EEPROM as nonvolatile memory.
The DAC interfaces with the base unit via I2C allow up to four MCP4728 to  connect and supports various I2C speeds such as Standard (100 kbps), Fast (400 kbps), and High Speed (HS) Mode (3.4 Mbps). The DAC output is between 0~-~3.3~V. However, the desired output range is \(\pm\) 4~V. The level-shifter is a non-inverted Texas Instruments OPA209, which has Op-amp gain of 2.42 and a 2.8 V bias voltage. It is used to shift and scale the signal from 0~-~3.3~V to \(\pm\) 4~V. The shifted and scaled output can be calculated as \inlineequation[eq:1]{Vo = 2.42*Vin - 4}. OPA209 is chosen due to it low power, low noise characteristics, and has a high short-circuit current rating. The MCP4728 has a resolution of 12 bits
together with the OPA209 level-shifter result in an output resolution of 1.95 mV

One feature that distinguishes the multi-channel potentiostat from other potentiostat designs to date is the ability to disconnect (open-circuit) individual output channels.  The analog switches are added to connect or disconnect individual electrodes, altering electrode configurations automatically. Maxim Integrated MAX326 is a quad CMOS Analog switch, Single Pole Single Throw (SPST), and Ultra-Low-Leakage. It allows the multi-channel potentiostat to connect or disconnect to the electrode independently. The analog switches have 10 pA maximum leak current and 2 pC typical charge injection. In a low current application, leak current from a switch could unintentionally actuate the electrochemical device making these ultra-low leak characteristics crucial.

The input stage of the multi-channel potentiostat measures the current from an adjustable voltage source that consists of an ammeter and Analog-to-Digital Converter (ADC). Unlike traditional potentiostat designs that use a transimpedance amplifier for current measurements, our multi-channel potentiostat has a shunt resistor. This approach might lower the accuracy but it allows simultaneous recording of multiple voltage sources. For instance, the multi-channel potentiostat with eight channels can operate four independent, two-electrode systems simultaneously, where four standard potentiostats would be required for the same setup. Additional, the benefits of the multi-channel potentiostat are unlike traditional potentiostats because it can interface with electrochemical cells with three or more electrodes.
The system is designed for high impedance load (above $1 M\Omega$) to measure $nA$ range currents.

The shunt resistors and instrumental op-amps are carefully chosen to minimize the error of the shunt resistor technique. INA121, a low-power instrumental amplifier with ultra-low bias current (4 pA), is used to mitigate measurement errors due to the bias current. The shunt resistor value is 10k 0.1\%, and the instrumental amplifier gain is 100. Consequently, the current to voltage conversion ratio is 1 nA to 1 mV. This design measures the current range from -1,650~to~+1,650~nA. The current measurement range is adjusted by changing the value of the gain resistor for the instrumental amplifier, or shunt resistor, or both. Due to the limited range of the ADC, the instrumental amplifier output (-1.65~to~+1.65~V) is shifted to match the 0~-~3.3~V of the ADC input. The output from the instrumental amplifier is fed into the inverted op-amp with an active low-pass filter to minimize noise and level-shifted by 1.65V. The inverted op-amp output is passed through another passive first-order low-pass filter with a cut-off frequency of 72 Hz to reduce the high-frequency noise. The OPA4209 is chosen as an op-amp for both input stages where it offers low voltage noise, quiescent current, offset voltage, and a short-circuit current of 65 mA. ADS1115 is a four single-ended or two differential inputs 16-bit ADC I2C interface with built-in programmable gain (PGA), low-drift voltage reference, and four pin-selectable addresses. This design used the ADS1115 as four single-ended inputs with a maximum sampling rate of 860 samples per second.
As a result, the input stage ranges from -1,650~to~+1,650~nA with resolution of 125~uV or 0.125~nA.

The standalone software consists of two main stages: initialization and execution. During the initialization stage, the script initializes all hardware components such as the DAC, ADC, shift-register, and real-time clock. Then, the calibration process starts by switching off analog switches (open-circuit), measuring ADC values during the no-load condition, and averaging the recordings to store as a baseline. After the calibration process, the program shows all predefined protocols that are stored in CSV format and prompt a user input to select which protocol to execute. Once the protocol is selected, the program will run until a user terminates it. While in the execution stage, the software displays real-time current measurement values. At the same time, it logs all of the states of the device, such as a timestamp, voltage output of different channels, and current measurement data in CSV format.

In addition to the predefined protocols that run in Standalone mode, the multi-channel potentiostat offers an Application Program Interface (API) where a user can control the multi-channel potentiostat remotely. The multi-channel potentiostat also connects with an external controller via an Ethernet or Wi-Fi connection. Similar to the standalone version, the API has the same initialization process. However, instead of executing a predefined protocol in the execution stage, it is listening for a User Datagram Protocol (UDP) command. The API allows users to set the voltage output of the multi-channel potentiostat and get current measurement feedback. Similar to standalone-mode, the multi-channel potentiostat states and output are also logged in CSV file format. API wrappers are available in both Python3 and Matlab libraries.

The PCBs are designed on a standard 1.6 mm FR4 two-layer PCB with a dimension of 72.5 x 57~mm (Base) and 71.7 x 57~mm (Stackable). Both PCBs are manufactured by PCBWay (China); the cost, including shipping is approximate ~\$ 2 per board. All electronic components are purchased from Digi-Key Electronics (MN, United States). The PCBs are populated and assembled at the University of California at Santa Cruz, shown in Figure \ref{fig:board_pic}. The total cost of the Base unit and Stackable board parts are respectively \$70 and \$110.

\section{Results and Discussions}
As an initial validation of the proposed device, we tested the linearity of the input and output signals. We also tested the performance of 8 channels multi-channel potentiostat under realistic load conditions. Figure \ref{fig:Device_Characteristic_a} and \ref{fig:Device_Characteristic_b} shows the measure output voltage as a function of the DAC output voltage and the error of the measured signal from a straight line. The output of the DAC is programmed to sweep between 0 and 3.3V, and it is measured using a digital multimeter. The result shows a linear relationship between the DAC's output and multi-channel potentiostat's output (Equation~\eqref{eq:1}) with a less than 0.5\% full-scale error.

Figures \ref{fig:Device_Characteristic_c} and \ref{fig:Device_Characteristic_d} show the Measured Input Voltage as a function of the Current and the error of the measured signal from a straight line. The output of the multi-channel potentiostat is connected to a $1 M\Omega$ resistor and sweeps between two voltages to generate a variety of current conditions. A digital multimeter is used to measure the output of an instrumental amplifier and level-shifting op-amp. Similarly, the output test results show the error from linearity (between the current and multi-channel potentiostat's ADC input) is less than 1\% of the full-scale.

Figures \ref{fig:Device_Characteristic_e} and \ref{fig:Device_Characteristic_f} show the
performance of the proposed device under a simulated experiment. The output of the multi-channel potentiostat is connected to a $1 M\Omega$ resistor to simulate a high-impedance load. The multi-channel potentiostat's output is programmed to sweep between -1.65 and +1.65 V, and the current is measured using the ADC. It is shown the current varies linearly with the voltage (between -1.65 $\mu A$ to +1.65 $\mu A$).
As additional validation of the proposed system, we calculated the load's resistance as 1.00969 M$\Omega$, compared to 0.997 M$\Omega$ when measured with a digital multimeter.

\begin{figure*}[ht]
    \centering
    \subfigure[The Multi-channel potentiostat output voltage vs. DAC voltage]{\label{fig:Device_Characteristic_a}\includegraphics[width=.44\textwidth]{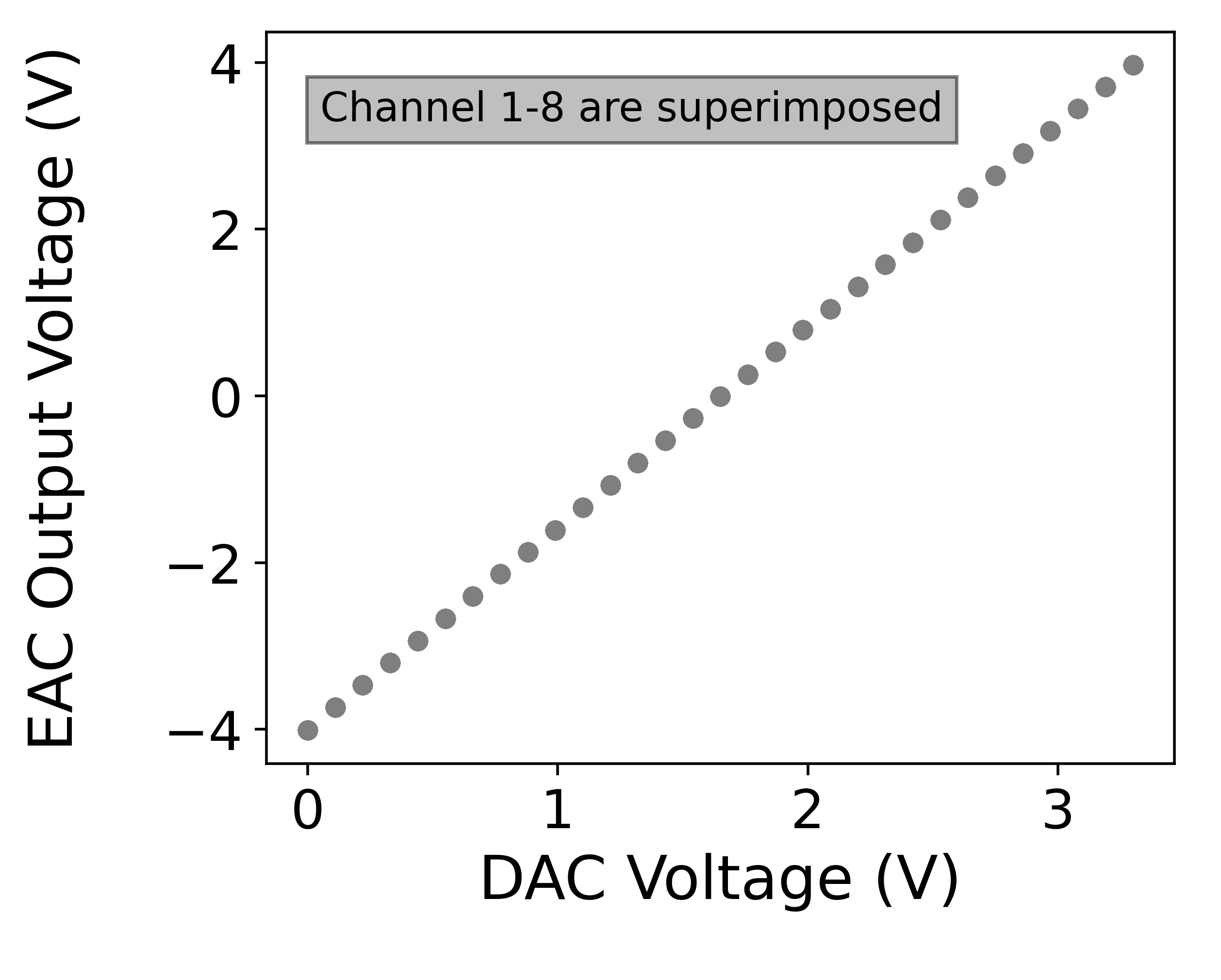}}
    \subfigure[Output voltage error]{\label{fig:Device_Characteristic_b}\includegraphics[width=.44\textwidth]{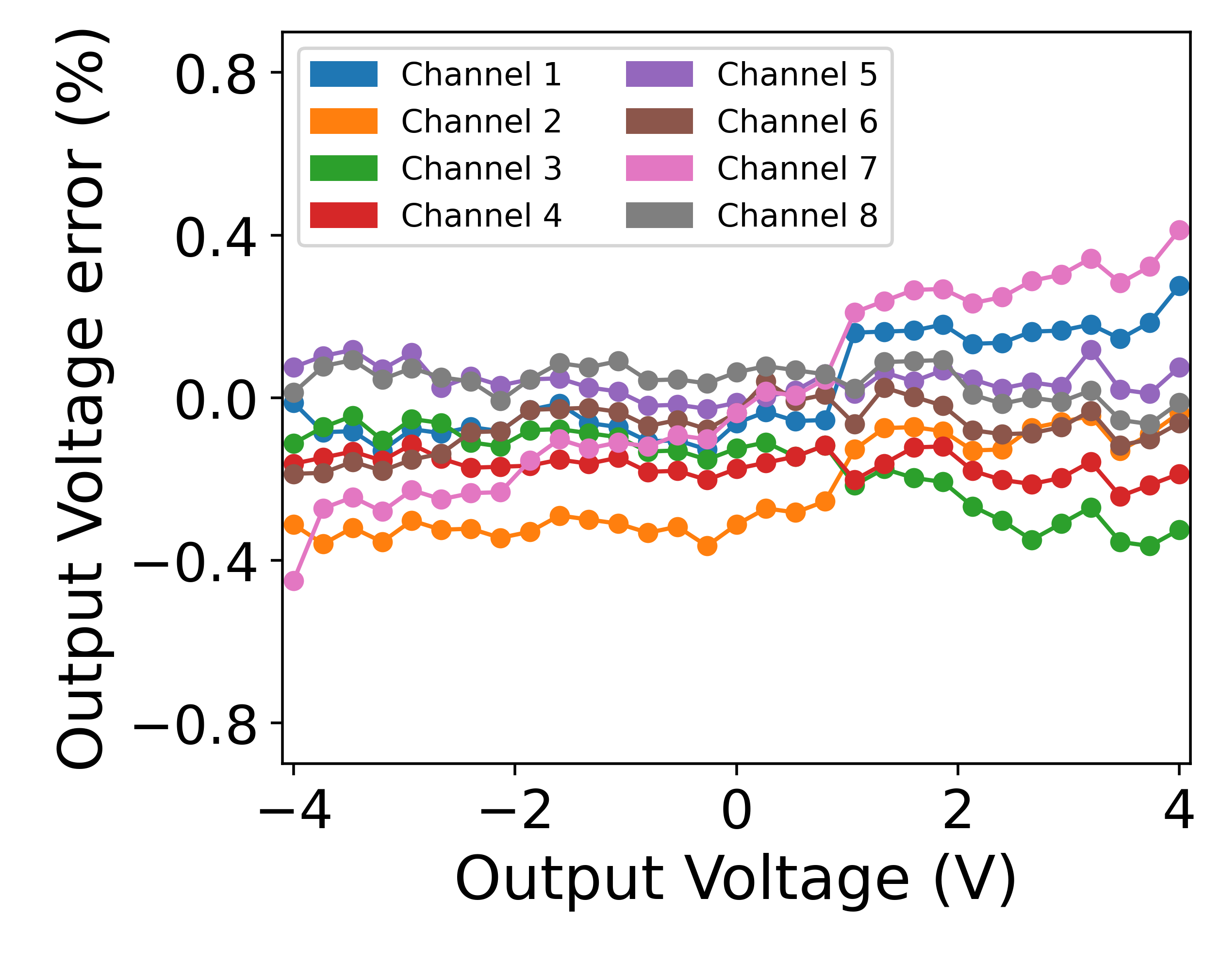}}
    \subfigure[Output voltage vs. input current for the ``current to voltage'' modules (Fig. \ref{fig:circuit_diagram})]{\label{fig:Device_Characteristic_c} \includegraphics[width=.44\textwidth]{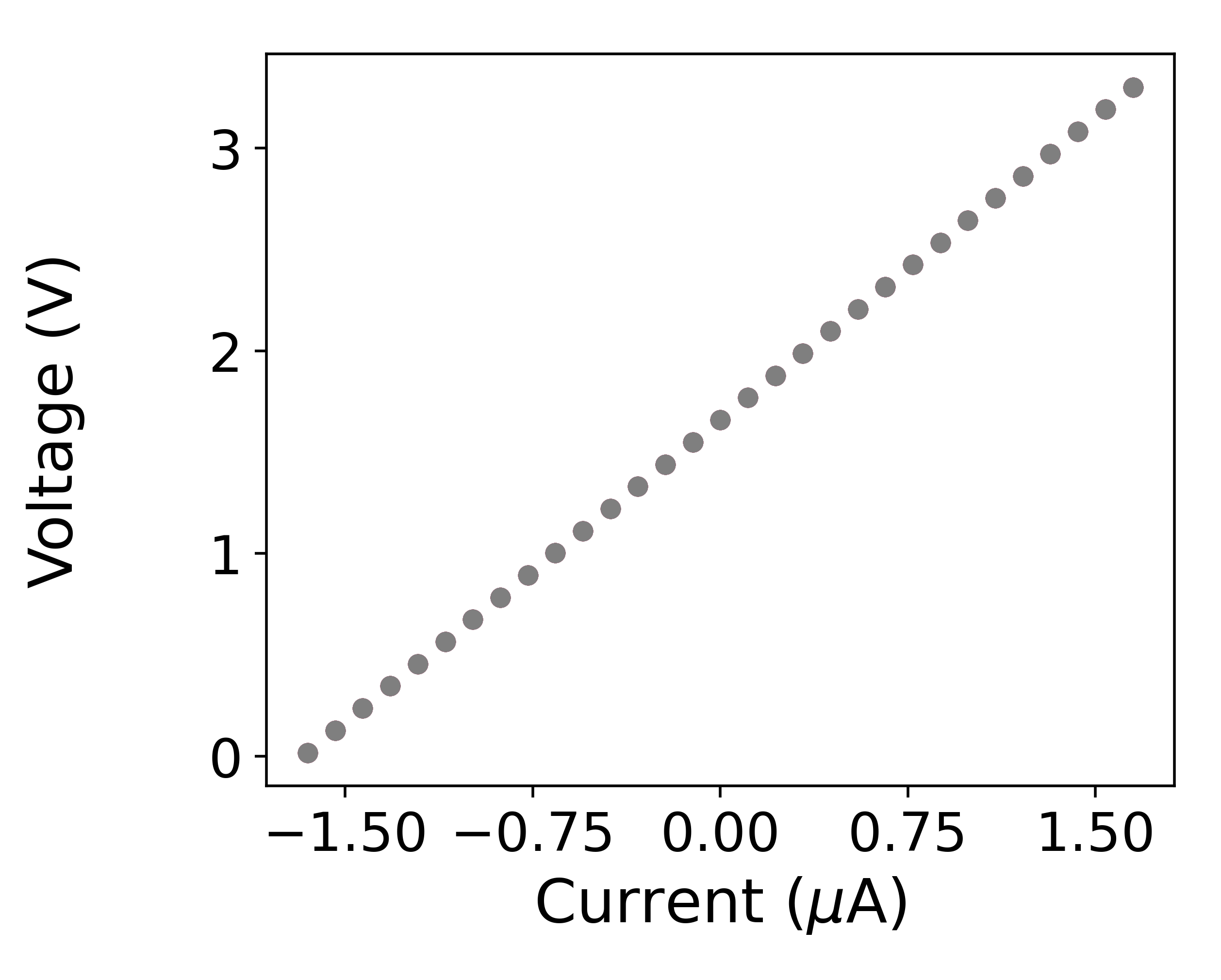}}
    \subfigure[``current to voltage'' modules input current error]{\label{fig:Device_Characteristic_d}\includegraphics[width=.44\textwidth]{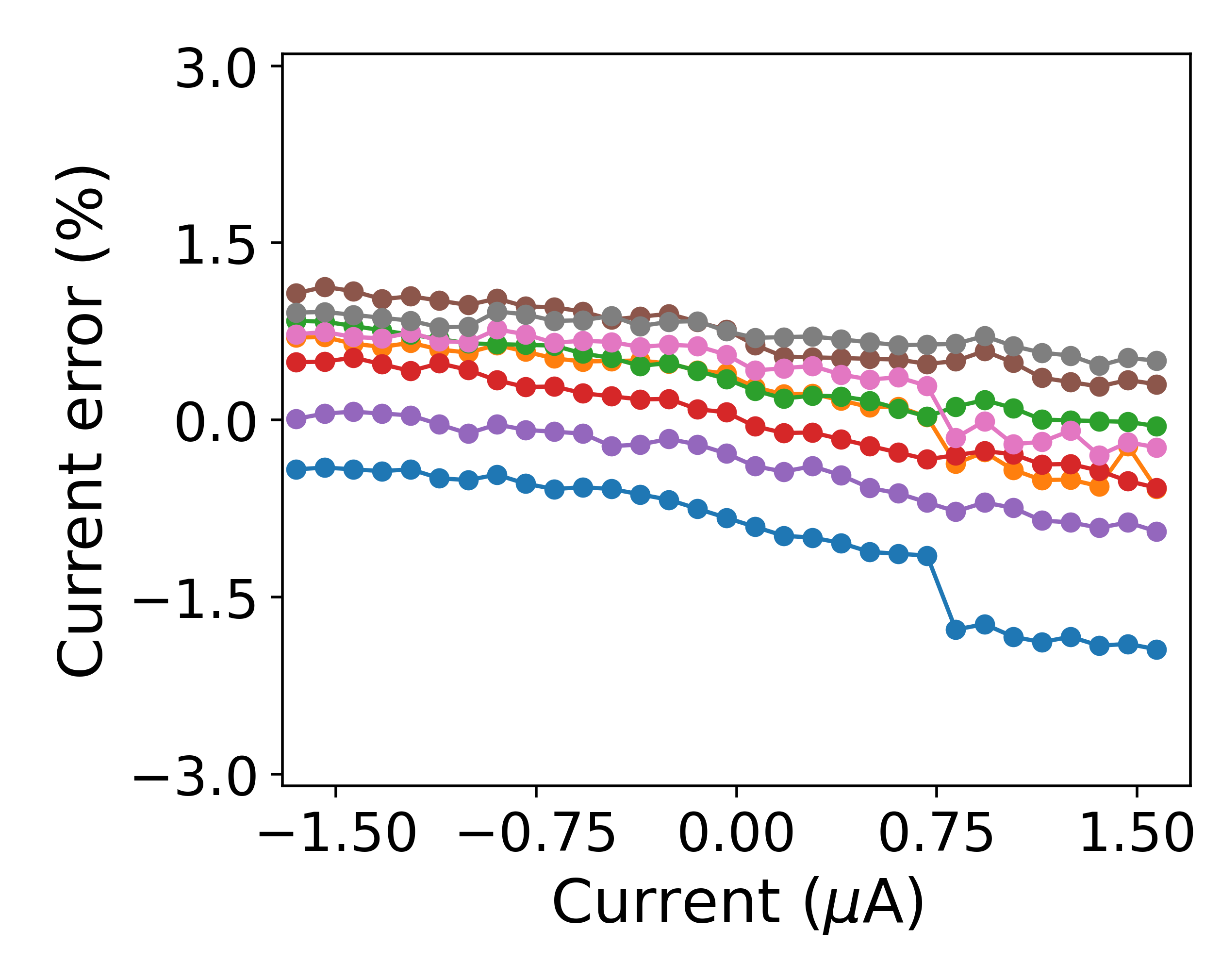}}
    \subfigure[The multi-channel potentiostat Load test using a $1 M\Omega$ resistor]{\label{fig:Device_Characteristic_e}\includegraphics[width=.44\textwidth]{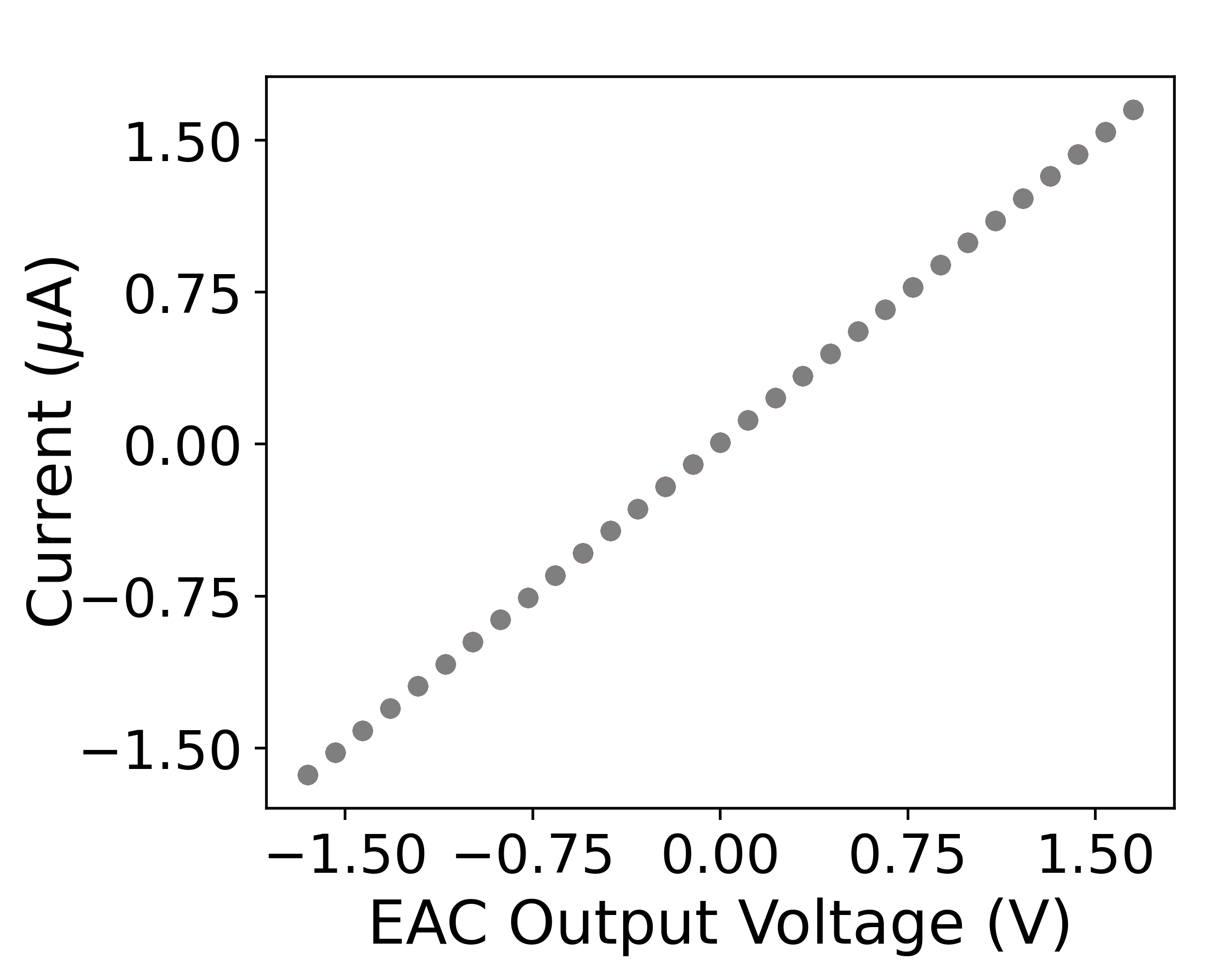}}
    \subfigure[The multi-channel potentiostat Load test errors]{\label{fig:Device_Characteristic_f}\includegraphics[width=.44\textwidth]{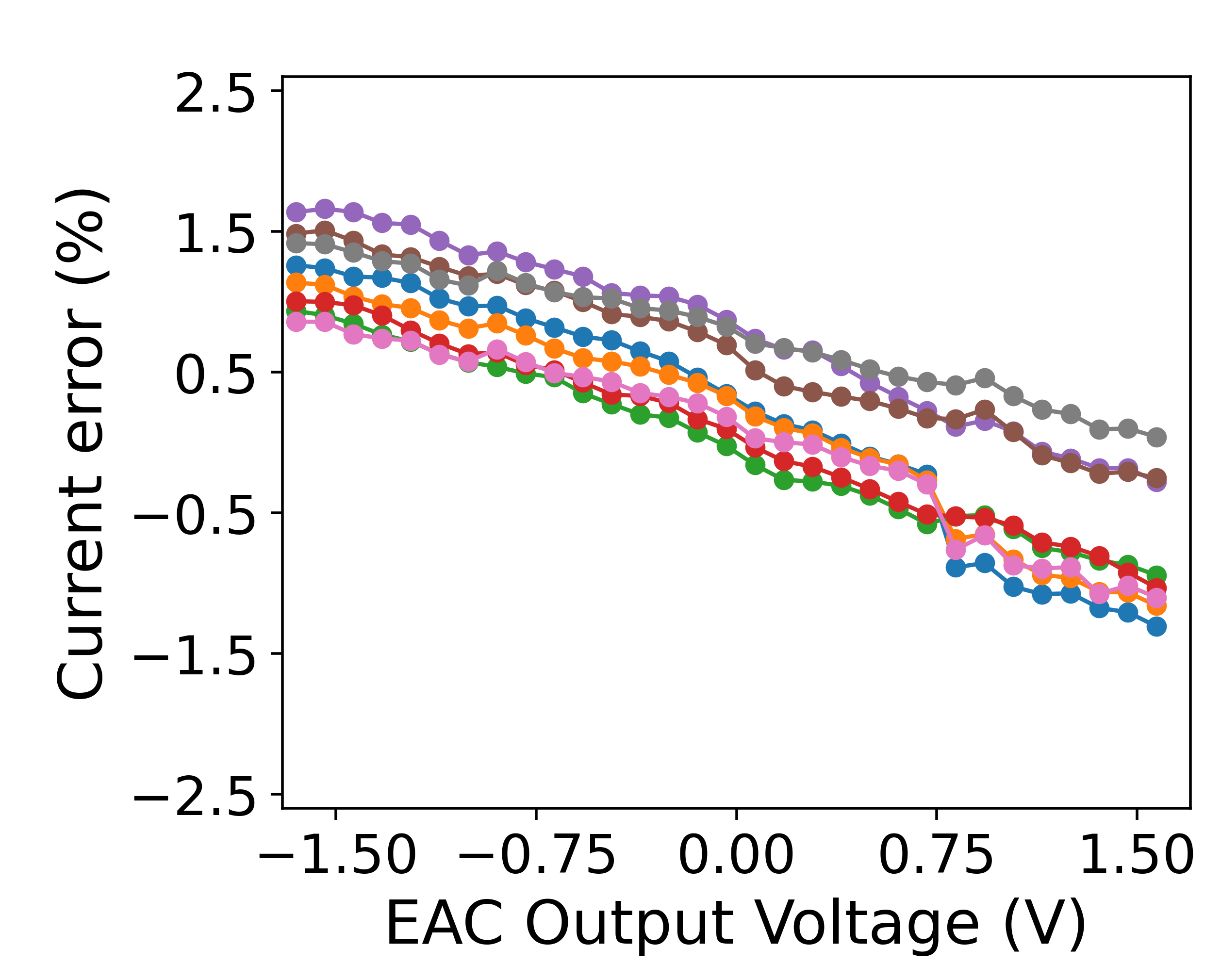}}
    \caption{The multi-channel potentiostat characterization tests for channels 1 to 8 (A) The measure output voltage on channels 1 to 8 (-4 to +4 V and a resolution of 1.95 mV) as a function of the DAC voltage. (B) Output voltage error. (C) The output voltage as a function of the input current (-1.65 to +1.65 $\mu A$) for the ``current to voltage'' converter modules shown in Fig. \ref{fig:circuit_diagram} (D) Input current errors on the ``current to voltage'' modules (E) multi-channel potentiostat Load test. The current passing through a $1 M\Omega$ resistor is monitored during a multi-channel potentiostat voltage sweep (-1.65 to +1.65 V) (F) Current errors during the load test.}
    \label{fig:Device_Characteristic}
\end{figure*}

\subsection{Cyclic Voltammetry Validation}
The multi-channel potentiostat’s performance during sensitive cyclic-voltammetry (CV) experiments is validated against a commercial system (Autolab PGSTAT128N). For both systems, we run linear sweep voltammetry tests between -0.9V and 0.5V. Figure \ref{fig:CV_setup} shows the schematic representation of the  experimental setup. The multi-channel potentiostat used a palladium nanoparticle coated gold (Au/PdNP) working electrode (250 $\mu m$  x 250 $\mu m$) and an Ag/AgCl pellet counter electrode  (2 $mm$ dia. x 2 $mm$ thick) in 50 $mM$ NaCl solution (pH 7). The palladium nanoparticle coating is chosen due to palladium's ability to reversibly form palladium hydride by absorbing hydrogen (H) when negatively biased in the presence of water \cite{grden2008electrochemical}. The experiment was conducted using three different scan speed 10, 50, and 100 mV/second.The 100mV scan rate shows the expected peaks for surface oxide formation (positive peak near 0.15 V), surface oxide reduction (negative peak near -0.2 V), hydrogen adsorption and absorption forming palladium hydride (peak near -0.65 V), and H desorption (positive peak near -0.75) shown in Figure \ref{fig:CV_100}.
The CV sweeps with this system are consistent with the PGSTAT128N over a range from 10 mV/s to 100 mV/s Figure \ref{fig:CV_10} - \ref{fig:CV_100}. The positive peak indicating H desorption is shifted when using the multi-channel potentiostat system by 50 mV. This is explained by the error of the multi-channel potentiostat's output stage and is improved through an additional voltage calibration step.

\begin{figure}[t]
    \centering
    \subfigure[NaCl Experiment setup]{\includegraphics[width=.4\textwidth]{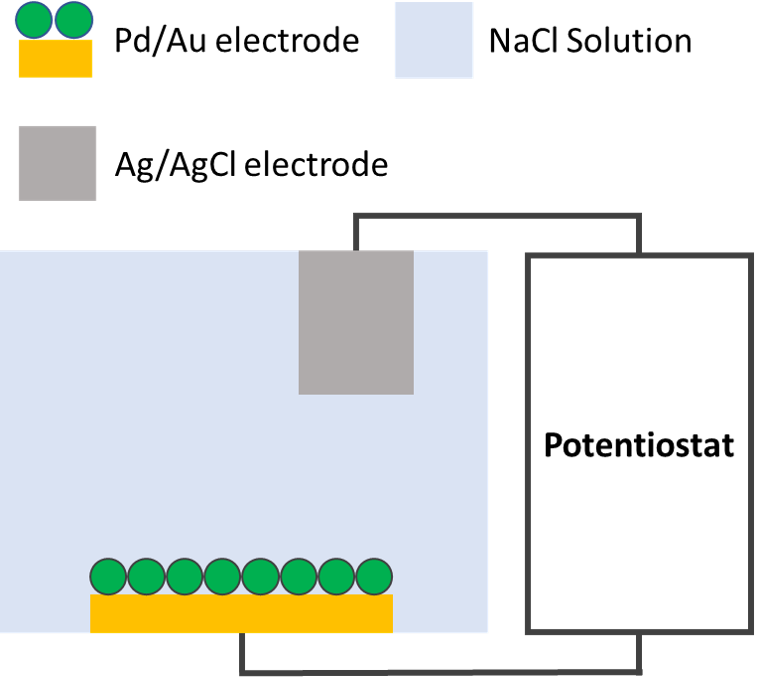}\label{fig:CV_setup}}
    \subfigure[CV 10 mV/s]{\includegraphics[width=.4\textwidth]{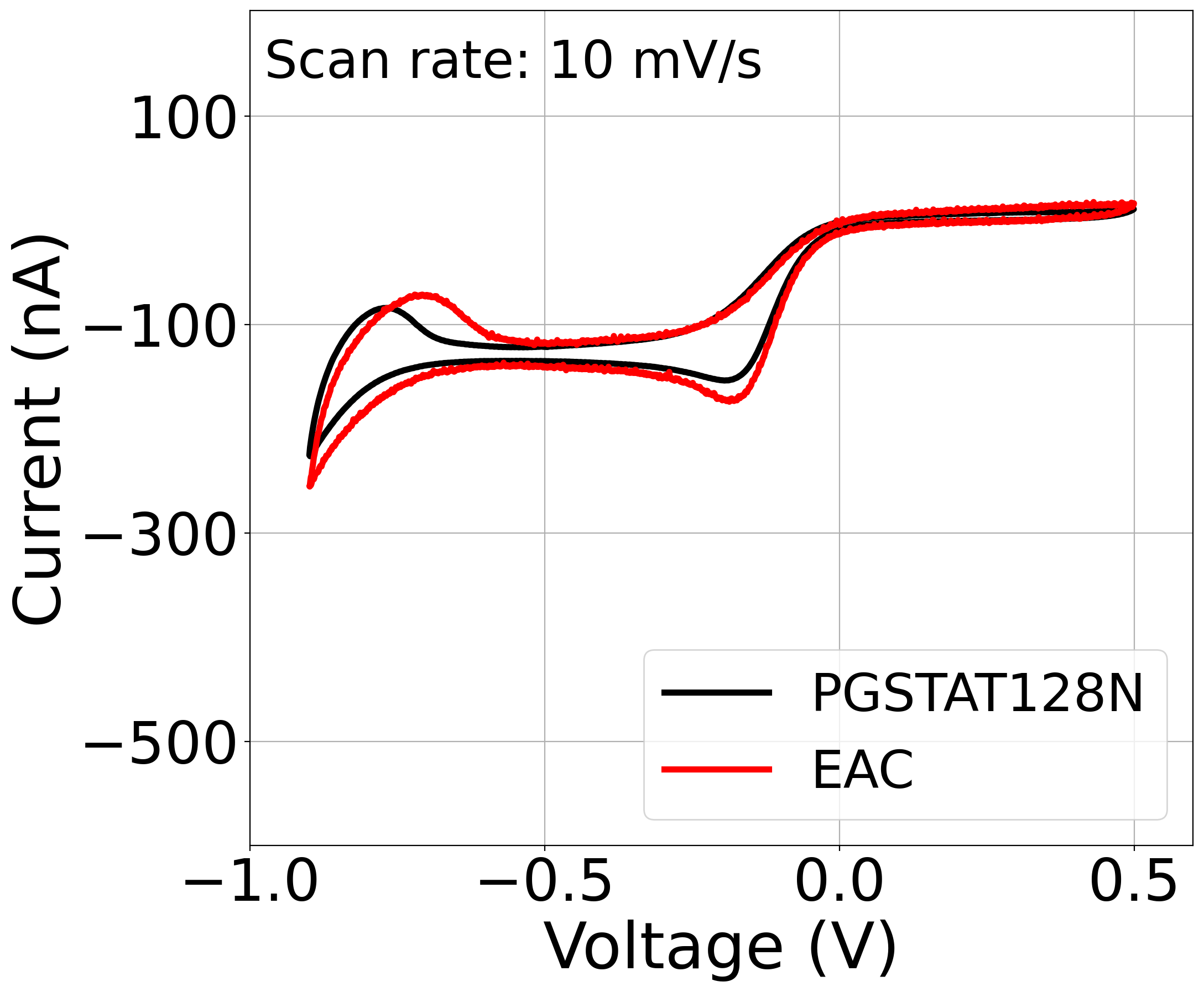}\label{fig:CV_10}}
    \subfigure[CV 50 mV/s]{\includegraphics[width=.4\textwidth]{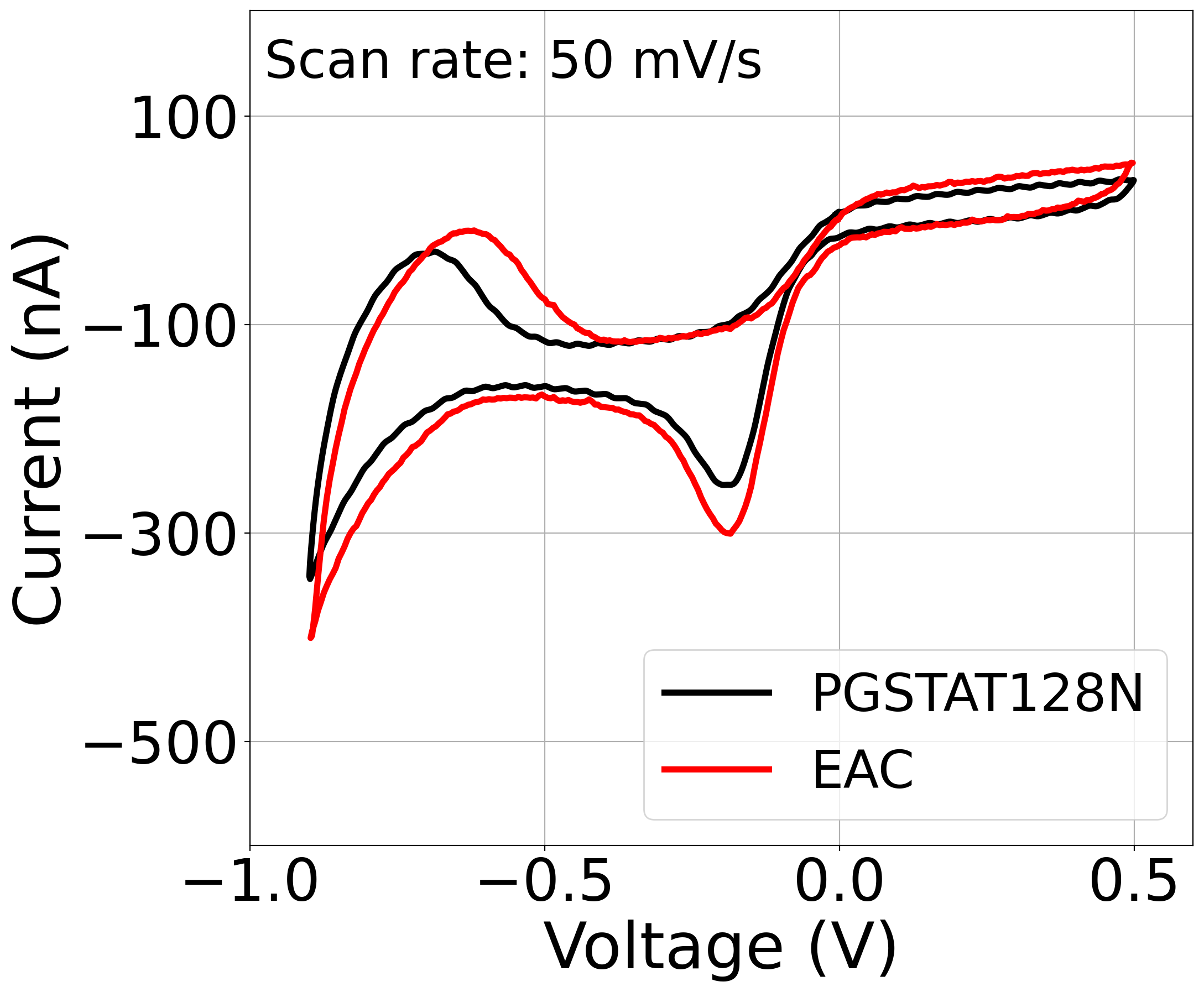}\label{fig:CV_50}}
    \subfigure[CV 100 mV/s]{\includegraphics[width=.4\textwidth]{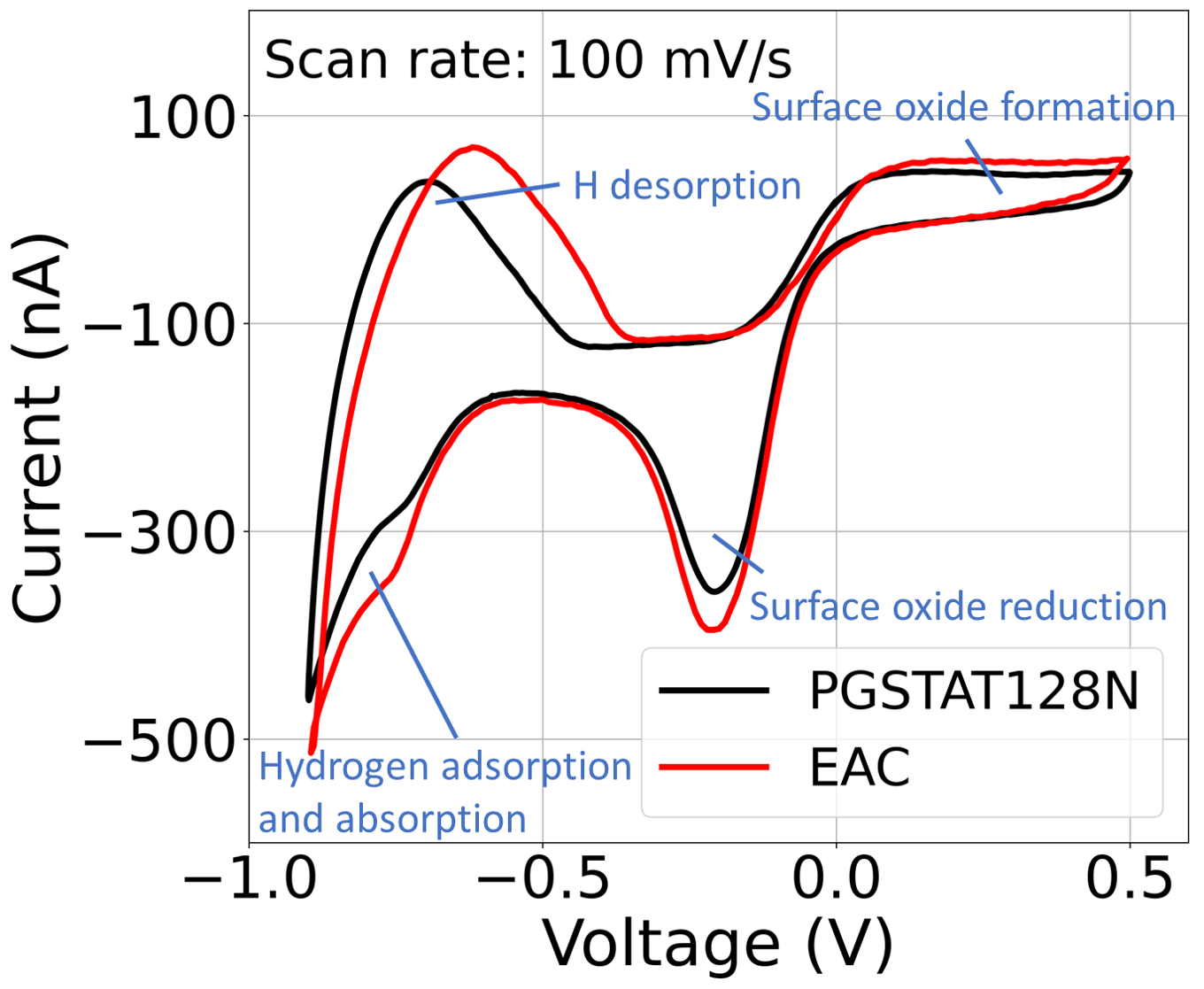}\label{fig:CV_100}}
    \caption{(A) Experimental CV setup with Pd functionalized Au electrode vs AgCl pellet electrode in NaCl solution. (B-D) CV of Pd NP/Au vs AgCl with various scan rates recorded with a commercial potentiostat and the multi-channel potentiostat device. (B) 10 mV/s scan rate (C) 50 mV/s scan rate (D) 100 mV/s scan rate.}
    \label{fig:CV2}
\end{figure}

\subsection{Ion pump Amperometry}
Amperometry is a technique used to study redox reactions by providing a DC voltage between two electrodes and measuring the generated current.
Ion pumps are bioelectronic devices that can manipulate physiological processes in vitro and in vivo delivering ions and small molecules directly to living matters \cite{selberg2020pump} \cite{jia2020clpump}. The devices operate by applying a DC voltage to the electrodes position in a reservoir and a target electrolyte, where ions are driven from the reservoir to the target through an ionic conducting material. Figure \ref{fig:amperometry:a} shows the experimental setup where the ion pump has 1 electrode in the reservoir and 9 individual electrodes in the target. The multi-channel potentiostat controls the ion pump and records the current with a 15Hz sampling rate. In contrast, the KEYENCE, BZ-X710 fluorescent microscope images the experiment using a 0.5Hz recording rate through a GFP filter. The multi-channel potentiostat is connected to nine working electrodes and one counter electrode of the ion pump. It is programmed to generate a square wave with amplitude +1.4 and -1.4V and a period of 30 seconds where it cycles from the first to the ninth electrode. While one electrode is active, the rest of the electrodes are on open-circuit mode.
Here we measure the proton concentration (pH) using fluorescence microscopy simultaneously while the voltage is swept. The target well is filled with pH sensitive fluorescent dye(SNARF\textregistered-1). In this process, each proton that moves from the reservoir to a certain electrode in the target results in an electron collected in the corresponding circuit channel.
Figure 6B shows the fluorescence intensity sampled from each electrode, which increases with fewer protons in the solution. By applying a negative voltage, the ion pump moves protons from the target to the reservoir resulting in higher fluorescence intensity and vice versa. The current profiles synchronize with the changes of fluorescence intensity in Figure \ref{fig:amperometry}C indicates that the multi-channel potentiostat can run amperometry experiments and provide reliable current data.
This experiment does not use electrodes 4 and 6 of the ion pump.
Overall, the amperometry results show good agreement with the actuation voltage and the fluorescent results.

\begin{figure}[t]
    \centering
    \subfigure[Spatial map of the Ion pump]{\includegraphics[width=.4\textwidth]{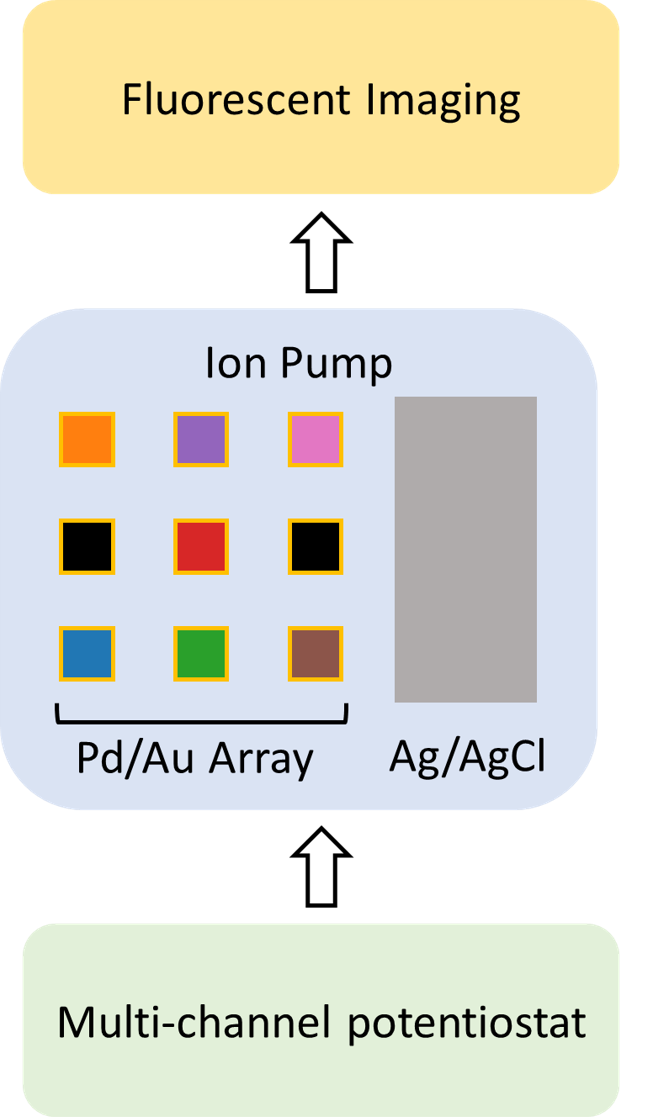}\label{fig:amperometry:a}}
    \subfigure[Fluorescent and Amperometry results]{\includegraphics[width=.5\textwidth]{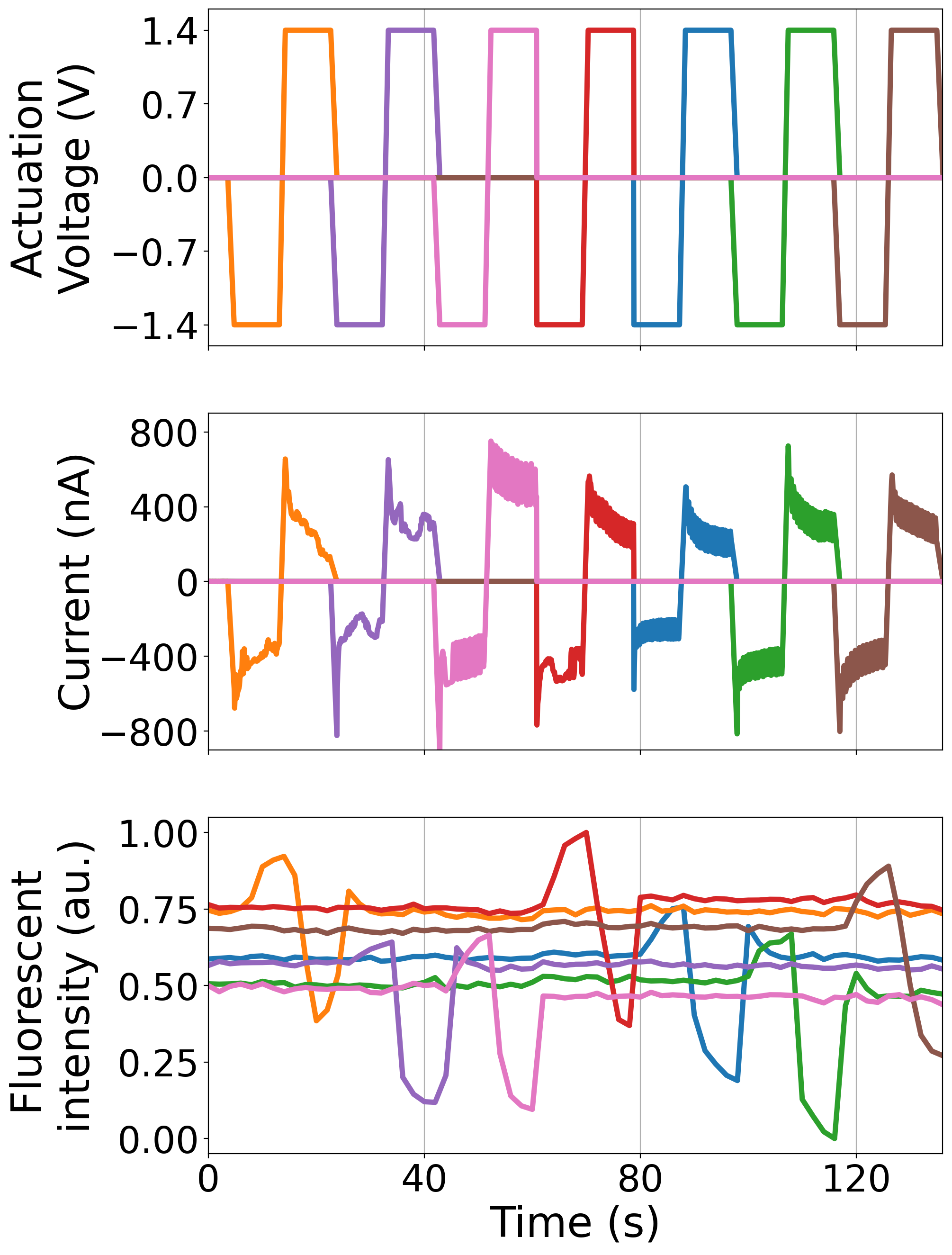}\label{fig:amperometry:b}}
    \caption{(A) Physical configuration of electrode placement of the ion pump. KEYENCE, BZ-X710 fluorescence microscope is used to image proton concentration (pH) every 2 seconds. Various colors represent different working electrodes and the black color indicates unused electrodes. (B) Actuation voltage and Amperometry results of in-house Ion pump and normalized fluorescent intensity correspond directly to $Cl^-$ ion concentration. The output shows that $Cl^-$ ion concentration changes according to the amperometry results from the ion pump.}
    \label{fig:amperometry}
\end{figure}

\subsection{Closed-loop control}
In many applications \cite{doi:10.1063/5.0027226}, instead of applying a predefined voltage (wave) in an open-loop fashion, the system needs to interact with the environment in real-time closed-loop form through feedback. To this end, there is a need for a versatile multi-channel potentiostat that allows control through custom algorithms. With the provided API, an external controller can remotely control the multi-channel potentiostat. These results were successfully demonstrated in the closed-loop feedback control experiment which is further detailed in \cite{jafari2020feedback}. In this work, Jafari et al \cite{jafari2020feedback} demonstrated this use-case in the following experiment. Similar to the amperometry experiment, this experiment utilizes the same ion pump device and monitors the changes in the ion concentration  using fluorescent microscope imaging. However, instead of a predefined square wave, the multi-channel potentiostat's output is generated using a higher lever control algorithm written as a Matlab script (external controller) running on a remote laptop. The Matlab script is running a machine learning-based control algorithm, in which the goal is to keep H+ ion concentration (fluorescent intensity) at a target value. In brief, the algorithm reads fluorescent images in real-time to further analyze and compute the mean value of the selected target area which is inversely proportional to the ion concentration. This value is fed into the machine learning-based control algorithm to generate the appropriate output and remotely send it to the multi-channel potentiostat to maintain the H+ ion concentration value to follow the desired target. The result shows that the control algorithm can successfully keep H+ ion concentration at the target function which shown in Figure \ref{fig:close_loop} which further demonstrates the seamless flexibility of the multi-channel potentiostat. It also allows an interface with an external controller running on a different machine.


\begin{figure}[h]
    \centering
    \includegraphics[width=\linewidth]{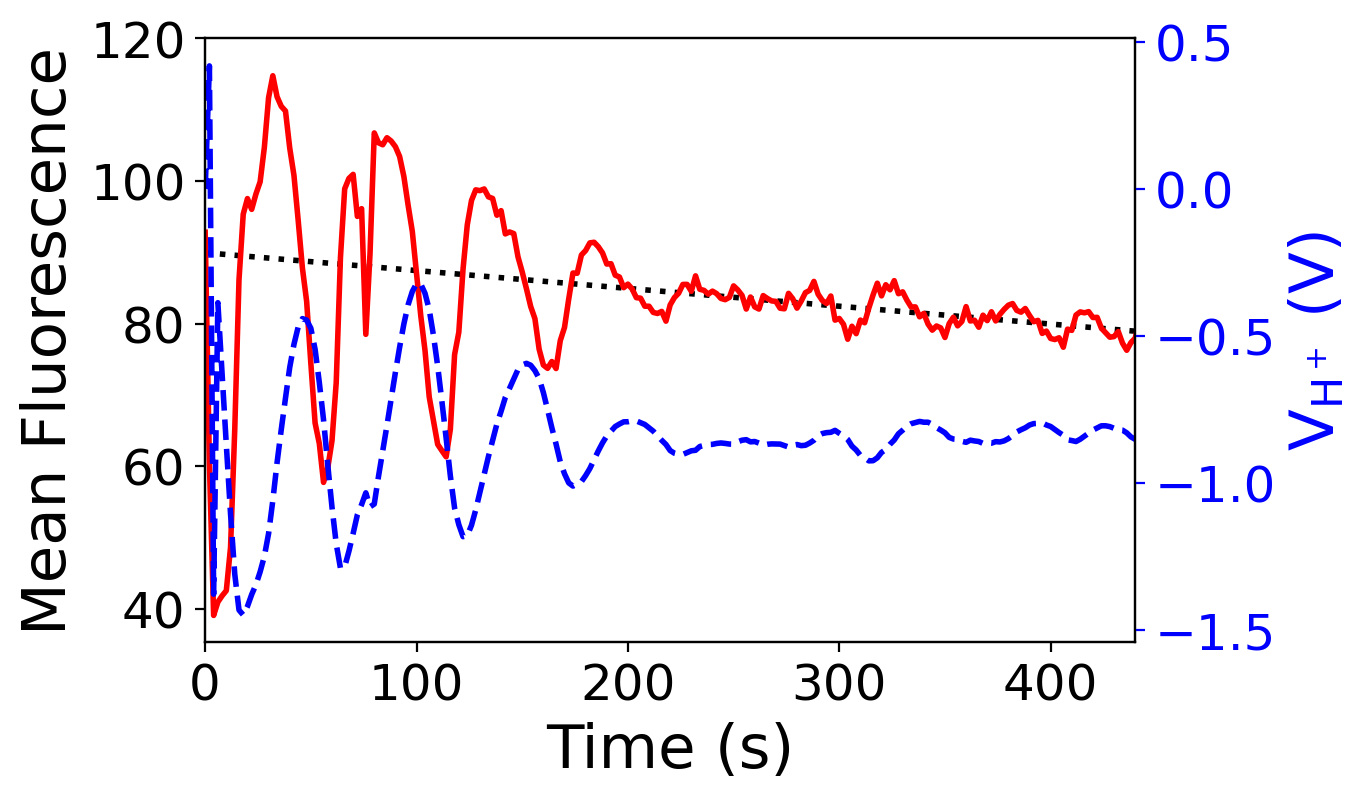}
    \caption{Close-loop control black line: Target; Red line: Fluoresce respond; and Blue: Controller output}
    \label{fig:close_loop}
\end{figure}

\section{Conclusion}
We have proposed a low-cost, multi-channel portable electrochemical array controller that allows users to conduct experiments with complex electrochemical cells (more than three-electrodes) in aqueous and non-aqueous solutions. The proposed design is modular: it is possible to scale up to 64 parallel potentiostats with adjustable voltage sources \(\pm\) 4 V and maximum currents \(\pm\) 1.5 $\mu A$.

The device is validated in a cyclic-voltammetry experiment against a commercial system (Autolab PGSTAT128N). To demonstrate the functionality of the multi-channel potentiostat device, we used it to control an ion pump that had one AgCl electrode in a NaCl solution and nine palladium individual target micro-electrodes. The amperometry results show good agreement with the actuation voltage and the fluorescent results.

The device could be remotely controlled via API enabling it to integrate with other laboratory systems. Here we had shown that it successfully integrated in an ion pump closed-loop controlled to monitor the changes into the ion concentration through imaging it using a fluorescent microscope.

Although the bench-top potentiostat would perform more proficiently in similar conditions, the multi-channel potentiostat is an ideal solution in scenarios that are not feasible for bench-top potentiostats, and a high degree of accuracy is not required (i.e., field-testing and point-of-care). In conclusion, we have proven that the proposed device is scalable, portable, significantly cheaper than commercially available systems with similar characteristics, and could be controlled remotely. Therefore it could be used as a component in a lab setting that needs to be operated remotely.


\begin{landscape}
\begin{table}
    \centering
    \begin{tabular}{|c+c|c|c|c|c|c|c|}
        \hline
        \multirow{3}{*}{Device} & Cyclic& Input & Output & Number & Size & \multirow{3}{*}{Interface} & \$ \\
        $ $ & Voltammetry& Range & Range & of & (mm) & $ $ & per\\
        $ $ &  $ $ & (nA) & (V) & Channel & $ $ & $ $ & channel\\
        \thickhline

        Multi-channel & \multirow{2}{*}{Yes} & \multirow{2}{*}{0.001 - 1.65} & \multirow{2}{*}{4} & \multirow{2}{*}{64} & \multirow{2}{*}{73 x 57} & \multirow{2}{*}{UDP} & \multirow{2}{*}{\$}\\
        potentiostat &  &  &  &  &  &  & \\ \hline
        CheapStat\cite{rowe2011cheapstat} & Yes & 0.1 - 50,000 & 0.99 & 1* & 140 x 66 & USART & \$\$\$\\ \hline
        M-P \cite{cruz2014low}  & Yes & 5 - 750 & 0.792 & 1* & - & - & \$\$\\ \hline
        KickStat \cite{hoilett2020kickstat} & Yes & 5 - 750 & 0.792 & 1* & 21 x 20 & - & \$\$\\ \hline
        Friedman et al \cite{friedman2012cost} & Yes & 200 - 2,000 & - & 1* & - & - & \$\$\$\$\\ \hline
        DStat \cite{dryden2015dsta} & Yes & - & 1.46 & 1* & 92 x 84 & USART & \$\$\$\\ \hline
        Gabriel N. Meloni \cite{meloni2016building} & Yes & 200,000 & 1 & 1* & - & - & \$\\ \hline
        Dobbelaere et al \cite{dobbelaere2017usb}  & Yes & 2 - 20,000 & 8 & 1* & 50 x 50 & USBSTACK & \$\$\$\\ \hline
        Miniature Glucose & \multirow{2}{*}{No} & \multirow{2}{*}{50} & \multirow{2}{*}{1} & \multirow{2}{*}{1*} & \multirow{2}{*}{28 x 25} & \multirow{2}{*}{USART} & \multirow{2}{*}{\$\$}\\
        measurement system \cite{adams2018miniature} &  &  &  &  &  &  & \\ \hline
        JUAMI \cite{li2018easily} & Yes & 10,000,000 & 2.5 & 1* & 87 x 75 & USART & \$\\ \hline
        UWED \cite{ainla2018open} & Yes & 180,000 & 1.5 & 1* & 80 x 40 & Bluetooth & \$\$\\ \hline
        MiniStat \cite{adams2019ministat} & Yes & 100 & 1.2 & 1* & 27 x 20 & USART & \$\$\\ \hline
        ABE-Stat \cite{jenkins2019abe} & Yes & - & 1.65 & 1* & 74 x 89 & Bluetooth & \$\$\\ \hline
        SweepStat \cite{glasscott2019sweepstat} & Yes & 15 - 1,500 & 1.5 & 1* & - & Labview & \$\$\\ \hline
        Emstat 3 \cite{Palmsens:2020}    & Yes & 0.001 - 10,000 & 3 & 1* & 67 x 50 & PSTrace** & \$\$\$\\ \hline
        Li et al \cite{li2004multi}   & Yes & - & 2.048 & 4* & Benchtop & Labview & \$\$\$\\ \hline
        Pseudopotentiostat \cite{cumyn2003design}  & Yes & - & 5 & 64* & Benchtop & - & \$\$\\ \hline
        PGSTAT128N \cite{MetrohmAutolab:2013}  & Yes & 0.01 - 1,000,000 & 10 & 4 & Benchtop & Autlab NOVA* & \$\$\$\$\$\\ \hline
        Arraystat-ULC \cite{Nuvan:2011}   & Yes & 0.003 - 10,000 & 15 & 25 & Benchtop & EZware** & \$\$\$\$\\ \hline
    \end{tabular}
\begin{flushleft}
    { Notes: 1 *: Number of working electrodes \hspace{1cm}2.**: Proprietary software}
\end{flushleft}
\caption{{\bf The multi-channel potentiostat in comparison with other potentiostat devices}}
\label{tab:comparison}
\end{table}
\end{landscape}

\section*{Acknowledgments}
This research is sponsored by the Defense Advanced Research Projects Agency (DARPA) through Cooperative Agreement D20AC00003 awarded by the U.S. Department of the Interior (DOI), Interior Business Center. The content of the information does not necessarily reflect the position or the policy of the Government, and no official endorsement should be inferred

\nolinenumbers

\bibliography{biba.bib}

\end{document}